\documentclass{applemlr}
\usepackage{amsmath}
\usepackage{enumerate}
\usepackage{algorithm}
\usepackage{algpseudocode}
\usepackage{amsfonts}
\usepackage{amsthm}
\usepackage{cleveref}
\usepackage{diagbox}
\usepackage{colortbl}
\usepackage{amssymb}
\usepackage{xspace}
\usepackage{wrapfig}
\usepackage{adjustbox}
\usepackage{tabularx}
\usepackage{booktabs}
\usepackage{mathtools}
\usepackage{tikz}
\usepackage{enumitem}
\usepackage{silence}
\usepackage{dsfont}
\usepackage[table]{xcolor}
\usepackage[dvipsnames]{xcolor}
\usepackage{multirow}
\usepackage{makecell}
\usepackage{xfakebold}
\usepackage{amsmath,amsfonts,bm}

\def\eqref#1{equation~\ref{#1}}
\def\1{\bm{1}}

\DeclareMathAlphabet{\mathsfit}{\encodingdefault}{\sfdefault}{m}{sl}
\SetMathAlphabet{\mathsfit}{bold}{\encodingdefault}{\sfdefault}{bx}{n}

\definecolor{textgray}{HTML}{6E6E73}
\usetikzlibrary{positioning, calc}
\usetikzlibrary{decorations.pathmorphing}

\makeatletter
\patchcmd{\wrong@fontshape}{\@gobbletwo}{}{}{}
\makeatother
\numberwithin{equation}{section}
\makeatletter
\AtBeginDocument{
  \urlstyle{sf}
  
}
\makeatother

\definecolor{light}{RGB}{125, 125, 125}
\crefname{tcb@cnt@pbox}{code}{code}
\Crefname{tcb@cnt@pbox}{Code}{Code}
\crefname{assumption}{assumption}{assumption}
\Crefname{assumption}{Assumption}{Assumptions}

\newtcolorbox[auto counter]{pbox}[2][]{
  colback=white,
  title=Code~\thetcbcounter: #2,
  #1,fonttitle=\sffamily,
  fontupper=\sffamily,
  arc=2pt,
  colframe=bgcolor,
  coltitle=fgcolor,
  colbacktitle=bgcolor,
  toptitle=0.25cm,
  bottomtitle=0.125cm
}

\makeatletter
\newcommand\applefootnote[1]{%
  \begingroup
  \renewcommand\thefootnote{}%
  \renewcommand\@makefntext[1]{\noindent##1}%
  \footnote{#1}%
  \addtocounter{footnote}{-1}%
  \endgroup
}
\makeatother

\definecolor{cverbbg}{gray}{0.90}

\usetikzlibrary{arrows.meta, positioning, calc, backgrounds, fit}

\definecolor{cEB}{HTML}{154360}
\definecolor{cEF}{HTML}{EAF2FB}
\definecolor{cTB}{HTML}{145A32}
\definecolor{cTF}{HTML}{EAFAF1}
\definecolor{cDB}{HTML}{641E16}
\definecolor{cDF}{HTML}{FDEDEC}
\definecolor{cFg}{HTML}{1C1C1C}
\definecolor{cPanelBg}{HTML}{F5F6FA}
\definecolor{cPanelBd}{HTML}{CDD0D9}
\definecolor{cHard}{HTML}{6C3483}
\definecolor{cHardF}{HTML}{F4ECF7}
\definecolor{cSoft}{HTML}{B9770E}
\definecolor{cSoftF}{HTML}{FDF2E1}
\definecolor{cEncW}{HTML}{D4EFDF}
\definecolor{cEncS}{HTML}{7DCEA0}
\definecolor{cEncL}{HTML}{196F3D}
\definecolor{cLMW}{HTML}{AED6F1}
\definecolor{cLML}{HTML}{1B4F72}

\tikzset{
  bigblock/.style={draw=cEB, fill=cEF, rounded corners=6pt, line width=1.6pt,
    minimum width=3.1cm, minimum height=2.5cm, align=center},
  smallblock/.style={draw=cTB, fill=cTF, rounded corners=6pt, line width=1.6pt,
    minimum width=3.1cm, minimum height=2.0cm, align=center},
  auxblock/.style={draw=cDB, fill=cDF, rounded corners=6pt, line width=1.3pt,
    minimum width=2.2cm, minimum height=0.95cm, align=center, font=\small},
  term/.style={draw=cFg!45, fill=white, rounded corners=4pt, line width=0.9pt,
    minimum width=1.9cm, minimum height=0.85cm, font=\small, align=center},
  opell/.style={draw=cFg!50, fill=gray!7, rounded corners=4pt, line width=0.9pt,
    minimum width=1.7cm, minimum height=0.75cm, font=\small, align=center},
  fwd/.style={-{Stealth[length=6pt, width=4.5pt]}, line width=1.3pt, draw=cFg},
  fwdlight/.style={-{Stealth[length=5pt, width=4pt]}, line width=0.85pt, draw=cFg!45, dashed},
  lbl/.style={font=\small, fill=white, inner sep=1.5pt, text=cFg}
}
\pgfdeclarelayer{background}
\pgfsetlayers{background,main}

\tikzset{
  blk/.style={draw=cFg!55, fill=white, rounded corners=3pt, line width=0.7pt,
    minimum height=8mm, minimum width=17mm, align=center,
    font=\sffamily\scriptsize, inner sep=3pt},
  tblk/.style={blk, draw=cEB, fill=cEF},
  sblk/.style={blk, draw=cTB, fill=cTF},
  iblk/.style={blk, draw=cFg!35, fill=gray!5},
  grp/.style={draw=cFg!25, rounded corners=5pt, line width=0.6pt,
    dash pattern=on 3pt off 2pt, inner sep=6pt},
  ar/.style={-{Stealth[length=4.5pt, width=3.4pt]}, line width=0.7pt, draw=cFg!75},
  tag/.style={font=\sffamily\scriptsize, text=cFg!70, inner sep=1.5pt},
  hdr/.style={font=\sffamily\bfseries\scriptsize, inner sep=2pt},
}
\tikzset{
  fslab/.style={draw=cEncL!85, fill=cEncS, rounded corners=1pt, line width=0.5pt,
    minimum width=#1, minimum height=3.8mm, rotate=90, align=center,
    font=\sffamily\tiny, inner sep=1pt, text=black},
  funit/.style={draw=black!55, fill=white, rounded corners=1.5pt, line width=0.6pt,
    align=center, font=\sffamily\scriptsize, inner sep=2pt},
  fghost/.style={funit, draw=black!30, fill=black!4, text=black!45,
    dash pattern=on 1.8pt off 1.4pt},
  fsig/.style={-{Stealth[length=5pt, width=3.6pt]}, line width=1pt, draw=black},
  fgrad/.style={-{Stealth[length=5pt, width=3.6pt]}, line width=0.9pt,
    draw=cTB, dash pattern=on 2.4pt off 1.8pt},
  fgname/.style={font=\sffamily\bfseries\small},
  fsym/.style={font=\sffamily\scriptsize},
}

\newcommand{\Rb}{\mathbb{R}}
\newcommand{\quant}{\mathcal{Q}}
\newcommand{\bridge}{\mathcal{G}}

\newcommand{\enc}{\mathcal{E}}        % encoder operator
\newcommand{\sub}{\mathcal{S}}        % subsampler
\newcommand{\conf}{\mathcal{C}}       % conformer stack
\newcommand{\down}{\mathcal{D}}       % post-downsample
\newcommand{\aff}{\mathcal{A}}       % affine width-matching layer
\newcommand{\Lmatch}{\mathcal{L}_{\mathrm{lat}}}
\newcommand{\Lnorm}{\mathcal{L}_{\mathrm{lat}}^{\mathrm{norm}}}

\title{Compressing Streaming Neural Audio Encoders via Latent-Space Distillation}

\author[*]{Prasanth Yadla}
\author[*]{Mohammad Samragh Razlighi}
\author{Dongseong Hwang}
\author{Mingbin Xu}
\author{Yuanyuan Zhang}
\author{Chung-Cheng Chiu}
\author[\dagger]{Yongqiang Wang}
\author[\dagger]{Yuan Liu}
\author{Zhen Huang}
\author{Xiaodan Zhuang}

\affiliation{Apple}

\contribution[*]{These authors contributed equally to this work. $^{\dagger}$Work done while at Apple.}

\abstract{
System-wide Dictation on Apple devices runs entirely on-device, and the speech it
transcribes reaches the foundation model through a tokenizer: an encoder that maps
short windows of waveform onto the representation the language model reads. Because
that model is sparsely activated under Instruction-Following Pruning, only a small
subset of its experts occupies DRAM at any time, so the always-on tokenizer
competes for the same memory, and its parameter count bears directly on power and
latency. In this work
we study how to compress such a tokenizer by distillation, taking as the
supervision target neither the discrete token nor the output distribution but the
\emph{pre-quantizer latent} the model actually consumes---the last representation
the two token interfaces share. We
train only the student encoder to regress the teacher's per-frame latent under a
squared-error objective, with a single affine layer absorbing the
teacher--student width mismatch. Because the target precedes both the quantizer
and the language-model bridge, one recipe covers both token interfaces we
support, and applies both to a tokenizer pretrained alone and to one jointly trained
with a language model. At $2.8\times$ compression the distilled student stays within $1.9\%$ relative
WER of its teacher on five of
six teacher--student pairs without any fine-tuning, and improves on an independently trained tokenizer of identical capacity by
$3.9\%$ relative.
}

\metadata[Keywords]{\sffamily audio tokenization, knowledge distillation, speech recognition, on-device inference}
\metadata[Correspondence]{\sffamily Prasanth Yadla: \url{p_yadla@apple.com}; Mohammad Samragh Razlighi: \url{m_samraghrazlighi@apple.com}}
\date{\sffamily\today}

\begin{document}

\maketitle

\section{Introduction}
\label{sec:intro}

The encoder described in this work serves as the audio front end for
system-wide Dictation, running entirely on-device alongside AFM~3 Core Advanced,
Apple's most powerful on-device foundation model~\citep{apple2026afm3}. AFM~3 Core Advanced is
natively multimodal and built on a sparse~\citep{fedus2022switchtransformersscalingtrillion},
20-billion-parameter architecture that
uses Instruction-Following Pruning
(IFP)~\citep{hou2025instructionfollowingpruninglargelanguage}: the full model is stored in flash memory
(NAND)~\citep{alizadeh2024llmflashefficientlarge}, and for each prompt a lightweight dense block
selects a small subset of input-dependent routed experts. These are reselected
periodically during generation, and they combine with a set of always-active
shared experts to form a dense model in DRAM, with 1--4 billion parameters
active at a time. Because only this
small active set occupies DRAM at any time, an always-on audio front end must run
within a tight memory and latency budget that it shares with the model it feeds.
Compressing the tokenizer enough to fit that budget is the problem we address.

Multimodal foundation models that accept speech do so through an audio
tokenizer: a learned encoder that converts a waveform into the sequence of
representations the language model consumes. In our system the tokenizer emits
one vector per 80\,ms of audio, and the language model either looks up a
quantized index of that vector in its embedding table or reads a projected
version of the vector directly. Either way, the tokenizer is on the critical
path for every audio frame the model ever sees.

On-device deployment imposes a fixed memory and compute budget that the tokenizer
and the language model must share. Because the tokenizer encoder is resident for
every frame, its footprint constrains the admissible size of the language model
directly, and therefore bounds attainable system quality prior to any consideration
of recognition accuracy. Our full-size encoder consumes a share of this budget that
leaves the language model underprovisioned. We consequently seek an encoder at
least $2.8\times$ smaller in parameters at no cost in recognition accuracy.

Training a compact encoder from scratch is feasible but forgoes available
structure. The teacher encoder already realizes a well-conditioned acoustic
representation, and, where the tokenizer has been trained jointly with a language
model, that language model has adapted to the geometry of the representation. Both
considerations favor distillation~\citep{hinton2015distillingknowledgeneuralnetwork} over independent
training.

The appropriate distillation target, however, is less immediate for a tokenizer
than for a conventional encoder, for two reasons. First, the encoder output reaches
the language model only through a quantizer or a bridge projection, so discrete
token identities are related to the latent by a non-differentiable $\arg\min$;
supervising them would require a relaxation or a straight-through
estimator~\citep{oord2018neuraldiscreterepresentationlearning,mentzer2023finitescalarquantizationvqvae}. Second, during tokenizer training
the encoder is attached to two auxiliary decoders, one for transcription and one
for audio reconstruction, whose combined parameter count is comparable to or larger
than that of the encoder; supervising their outputs would allocate student capacity
to components the on-device tokenizer does not contain. We therefore define the objective
on the pre-quantizer latent, the final representation common to both token
interfaces and the quantity on which downstream accuracy depends. This target is
motivated by construction rather than established by ablation against the
alternatives.

Concretely, we train the student encoder alone to regress the teacher's per-frame
latent under a squared-error objective, with a single affine layer reconciling the
width mismatch between teacher and student. Because the target precedes both the
quantizer and the bridge, one recipe applies to either token interface; because it
is defined on the encoder, neither decoder is instantiated for the student. We apply
the recipe at two points in the teacher's lifecycle: after the tokenizer has been
pretrained in isolation, and after it has been jointly trained with a language
model.

At $2.8\times$ compression or more, the resulting student degrades by at most
$1.9\%$ relative WER against its teacher on five of six teacher--student pairs, in
each case without student fine-tuning. These pairs span both distillation stages,
both token interfaces, and three joint models. The student additionally outperforms
an independently trained tokenizer of identical capacity by $3.9\%$ relative, which
separates the contribution of the teacher signal from that of the reduced
architecture. Throughout we report relative differences, since an absolute WER gap
is uninterpretable without its baseline. Controlled comparisons then characterize
the influence of distillation stage, token interface, and teacher quality on student
accuracy.

\section{Related Work}
\label{sec:related}

Learned discrete latents originate with VQ-VAE~\citep{oord2018neuraldiscreterepresentationlearning}, and neural
audio codecs carried the idea to waveforms: SoundStream~\citep{zeghidour2021soundstreamendtoendneuralaudio},
EnCodec~\citep{défossez2022highfidelityneuralaudio} and
DAC~\citep{kumar2023highfidelityaudiocompressionimproved} established residual vector
quantization~\citep{lee2022autoregressiveimagegenerationusing} as the standard way
to turn a continuous encoder output into a compact sequence of integers, and the
token inventories they produce are what allow a language model to consume audio
directly, as in AudioLM~\citep{borsos2023audiolmlanguagemodelingapproach} and Moshi~\citep{défossez2024moshispeechtextfoundationmodel}.
A parallel line derives discrete units for recognition rather than reconstruction,
beginning with vq-wav2vec~\citep{baevski2020vqwav2vecselfsupervisedlearningdiscrete} and
w2v-BERT~\citep{chung2021w2vbertcombiningcontrastivelearning}. Closest to our objective are tokenizers that are
themselves distilled: SpeechTokenizer~\citep{zhang2024speechtokenizerunifiedspeechtokenizer} aligns the
first residual quantization level with HuBERT representations, and
RepCodec~\citep{huang2024repcodecspeechrepresentationcodec} learns a codec over speech representations rather
than waveforms. Both distill \emph{into} the tokenizer latent, as we do, but to
change what the tokens mean; we distill to shrink the encoder while holding the
token inventory fixed.

Both interfaces of Section~\ref{sec:interface} are established practice rather than
local conventions. AudioPaLM~\citep{rubenstein2023audiopalmlargelanguagemodel} places discrete audio
tokens directly in the language model's vocabulary, which is the discrete-token
interface. The continuous interface---a projection from encoder output into the
embedding space of
an otherwise unmodified language model---is the arrangement
of~\citep{fathullah2023promptinglargelanguagemodels} and of SLAM-ASR~\citep{ma2024embarrassinglysimpleapproachllm}, the latter
showing that a linear projector suffices for strong recognition, which is the
evidence behind our choice of a linear bridge. Qwen2-Audio~\citep{chu2024qwen2audiotechnicalreport}
and SALMONN~\citep{tang2024salmonngenerichearingabilities} adopt the same continuous interface for general
audio-conditioned models. Because both routes are in wide use, an objective placed
before either is portable beyond our system.

Distillation transfers behavior from a large model to a small one by supervising
the student on the teacher's output~\citep{hinton2015distillingknowledgeneuralnetwork}. For sequence
models the default expectation is sequence-level distillation on teacher
hypotheses~\citep{kim2016sequencelevelknowledgedistillation}, which we do not use: our objective needs no decoding
and no labels. Since an encoder emits a representation rather than a decision, the
productive variants match intermediate features in the manner of
FitNets~\citep{romero2015fitnetshintsdeepnets,Gou_2021}, and in language modeling
TinyBERT~\citep{jiao2020tinybertdistillingbertnatural} and MiniLM~\citep{wang2020minilmdeepselfattentiondistillation} regress hidden
states across a width mismatch using exactly the affine layer we adopt in
Section~\ref{sec:student}. For speech, DistilHuBERT~\citep{chang2022distilhubertspeechrepresentationlearning} and
FitHuBERT~\citep{lee2022fithubertgoingthinnerdeeper} compress self-supervised encoders such as
wav2vec~2.0~\citep{baevski2020wav2vec20frameworkselfsupervised} and HuBERT~\citep{hsu2021hubertselfsupervisedspeechrepresentation}, reporting
that hidden-layer matching transfers more signal than output matching alone. Two
strands of that literature bear directly on our ablations: the teacher--student
capacity gap, where a stronger teacher is not always a better teacher for a small
student~\citep{cho2019efficacyknowledgedistillation,mirzadeh2019improvedknowledgedistillationteacher}, and the separation between fidelity
to the teacher and generalization~\citep{stanton2021doesknowledgedistillationreally}, which is the distinction
our stage-1 results turn on.

The nearest compression baselines combine distillation with structured pruning, as
in DPHuBERT~\citep{peng2023dphubertjointdistillationpruning}, or train elastic supernets over width and depth,
as in LightHuBERT~\citep{wang2022lighthubertlightweightconfigurablespeech}; PARP~\citep{lai2021parppruneadjustreprune} prunes without
distilling. Distil-Whisper~\citep{gandhi2023distilwhisperrobustknowledgedistillation} compresses
Whisper~\citep{radford2022robustspeechrecognitionlargescale} by copying layers and training on pseudo-labels,
which differs from our setting in two respects: it requires transcripts, and it
distills the decoder we discard. On the architectural side,
Squeezeformer~\citep{kim2022squeezeformerefficienttransformerautomatic} and Zipformer~\citep{yao2024zipformerfasterbetterencoder}
reallocate capacity across depth and temporal resolution in Conformer encoders,
which is the same trade we exploit in Section~\ref{sec:student}. What distinguishes
our setting is the output boundary: prior work compresses either a self-supervised
encoder with no designated output, or a recognition model end to end using labels,
whereas the encoder here has its output fixed by a downstream quantizer or bridge,
and needs no labels at all.

\section{Distilled Tokenizer}
\label{sec:model}

Our distilled tokenizer is a student encoder trained to reproduce the per-frame
latent of a large teacher encoder. It is paired with the interface block that
connects that latent to the language model, either a quantizer or a bridge
projection. The distinctive feature of our approach is the choice of supervision
target: the pre-quantizer latent, rather than the discrete token or the output
of any decoder head.

\begin{figure}[t]
\centering
\resizebox{\textwidth}{!}{%
\begin{tikzpicture}[x=1mm, y=1mm]

  %% ── waveform ────────────────────────────────────────────────────────
  \foreach \i/\h in {0/2.1,1/4.8,2/3.1,3/6.7,4/4.2,5/8.0,6/3.3,7/5.9,8/2.5,
                     9/6.3,10/3.7,11/7.3,12/2.9,13/5.0,14/2.1}
    \draw[line width=1pt, draw=black!32] (\i*1.2,-\h) -- (\i*1.2,\h);

  \draw[fsig] (19.5,0) -- (24.5,0);
  \node[funit, minimum width=6mm, minimum height=21mm] (fb) at (28,0)
        {\rotatebox{90}{filterbank}};
  \node[fsym, text=black!55, anchor=north] at (28,-12) {$10$\,ms};

  %% ── encoder: the wedge taper is the 8x temporal reduction ───────────
  \draw[fsig] (31.2,0) -- (35.6,0);
  \fill[cEncW] (37,-13) -- (90,-6.5) -- (90,6.5) -- (37,13) -- cycle;
  \draw[draw=cEncL!40, line width=0.7pt]
    (37,-13) -- (90,-6.5) -- (90,6.5) -- (37,13) -- cycle;

  \node[fslab={20.6mm}] at (42,0) {subsampler\\$4\times$};
  \foreach \x/\hh in {51/18.6mm, 57/17.4mm, 63/16.2mm}
    \node[fslab=\hh] at (\x,0) {Conformer block};
  \node[font=\sffamily\tiny, text=cEncL, rotate=90] at (67.4,0) {$\times\,n$};
  \node[fslab={13.8mm}] at (73,0) {post-downsample\\$2\times$};
  \node[fslab={11.6mm}, dash pattern=on 1.6pt off 1.2pt] at (83,0)
        {affine\\$d_S\!\to\!d_T$};

  \node[fgname, text=cEncL, anchor=north] at (63.5,-16) {Encoder};

  %% ── pre-quantizer latent ────────────────────────────────────────────
  \draw[fsig] (90,0) -- (96,0);
  \foreach \i in {0,...,7}
    \draw[draw=cDB!70, fill=cDF, line width=0.4pt]
      (97.4,-6+\i*1.5) rectangle (100.4,-4.7+\i*1.5);
  \node[fsym, text=cDB, anchor=north] at (98.9,-7.2) {$\Rb^{T\times d_T}$};

  %% the objective, routed to a symbol at the figure edge
  \draw[-{Stealth[length=4pt, width=3pt]}, line width=0.6pt, draw=cDB]
    (98.9,6.6) -- (98.9,20) -- (147,20);
  \node[font=\sffamily, text=cDB, anchor=west] at (148.2,20)
        {$\mathcal{L}_{\mathrm{lat}}$};

  %% ── the two interfaces ──────────────────────────────────────────────
  \draw[fsig, draw=cHard] (100.8,1.3) -- (105,1.3) -- (105,9.5) -- (110,9.5);
  \draw[fsig, draw=cSoft] (100.8,-1.3) -- (105,-1.3) -- (105,-9.5) -- (110,-9.5);

  \node[funit, draw=cHard, fill=cHardF, minimum width=16mm, minimum height=7mm]
        (q) at (119,9.5) {quantizer $\quant$};
  \node[funit, draw=cSoft, fill=cSoftF, minimum width=16mm, minimum height=7mm]
        (g) at (119,-9.5) {bridge $\bridge$};

  %% discrete indices out of the quantizer
  \draw[fsig, draw=cHard] (127.4,9.5) -- (131.4,9.5);
  \foreach \i in {0,...,3}
    \draw[draw=cHard!85, fill=cHardF, line width=0.45pt]
      (132.6+\i*3.2,8.1) rectangle (135.1+\i*3.2,10.9);
  %% a continuous vector out of the bridge
  \draw[fsig, draw=cSoft] (127.4,-9.5) -- (131.4,-9.5);
  \draw[draw=cSoft!85, fill=cSoftF, line width=0.5pt, rounded corners=0.8pt]
    (132.6,-10.9) rectangle (145.1,-8.1);
  \foreach \i in {1,...,7}
    \draw[draw=cSoft!30, line width=0.3pt]
      (132.6+\i*1.56,-10.9) -- (132.6+\i*1.56,-8.1);

  \node[fgname, text=cHard, anchor=south] at (128,12.8) {discrete tokens};
  \node[fgname, text=cSoft, anchor=north] at (128,-12.8) {continuous embeddings};
  \node[fgname, text=black!55, anchor=north] at (119,-16) {Interface};

  %% ── language model ──────────────────────────────────────────────────
  \node[funit, draw=cLML, fill=cLMW, minimum width=12mm, minimum height=27mm,
        line width=0.8pt] (lm) at (163,0) {\rotatebox{90}{language model}};
  \draw[fsig, draw=cHard] (147.7,9.5)  -- (152,9.5)  -- (152,3.2) -- (156.6,3.2);
  \draw[fsig, draw=cSoft] (147.7,-9.5) -- (152,-9.5) -- (152,-3.2) -- (156.6,-3.2);
\end{tikzpicture}}
\caption{The tokenizer signal path and the two interfaces it feeds. The encoder is
the only part that is distilled and the only part that runs on-device; its wedge
narrows to indicate the $8\times$ temporal reduction that turns one filterbank
frame per $10$\,ms into one output vector per $80$\,ms. The affine layer is drawn with a
broken outline because it exists only while distilling. The quantizer and the
bridge are inherited from the teacher and are never trained: the discrete route emits
integer indices, drawn as separate cells, which the language model looks up in its
embedding table, while the continuous route emits a projected vector, drawn as a
continuous bar, which the model consumes directly. Both branch \emph{after} the
encoder and read the same pre-quantizer latent, which is why the single objective
$\mathcal{L}_{\mathrm{lat}}$ of Figure~\ref{fig:distill}, attached at that point,
serves both.}
\label{fig:interfaces}
\end{figure}
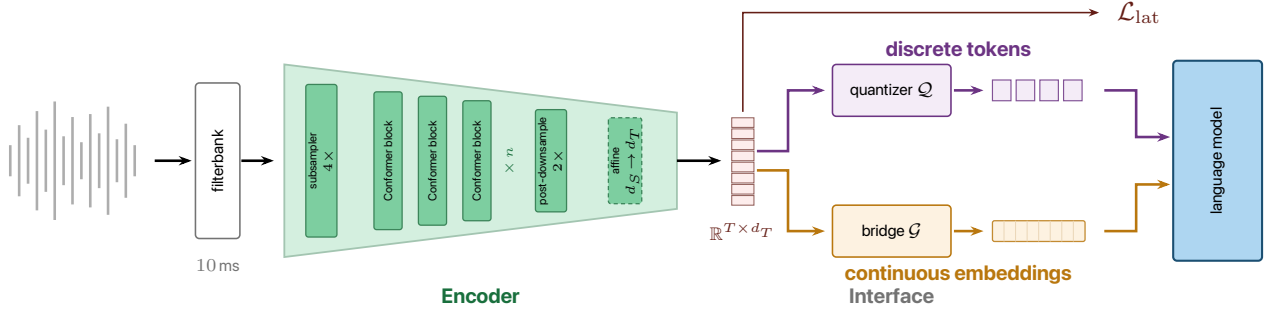

\subsection{Tokenizer--Model Interface}
\label{sec:interface}

The tokenizer contains an internal encoder block that converts every 80\,ms of
audio into a vector $h_t \in \mathbb{R}^{d_T}$. Two mechanisms connect that
vector to the language model, and the choice affects what must be carried
alongside the encoder but not the distillation objective itself.

Under the first, the discrete-token interface, the vector is mapped to the index
of its nearest codebook entry, $r_t=\arg\min_{c}\lVert h_t-e_c\rVert_2$,
and the language model treats that index as an entry in its word embedding table.
This keeps the input interface identical to text and adds no trained parameters at
the boundary. Under the second, the continuous interface, the vector is instead mapped to the
language model's embedding dimensionality by a linear \emph{bridge}
layer, and the model consumes the result directly, bypassing the word embedding
table:
\begin{align}
  u_t &= W_{\text{bridge}}\, h_t , \qquad W_{\text{bridge}} \in \mathbb{R}^{d_{\text{LM}} \times d_T} .
\end{align}
The continuous interface avoids the information loss of quantization at the cost of a
tokenizer-specific parameter block that lives logically with the language model
and must be carried along whenever the tokenizer is extracted from a joint
checkpoint. The two are referred to internally as hard and soft tokens; we use the
discrete and continuous naming throughout.

Figure~\ref{fig:interfaces} shows the branch point. Both interfaces read from the
same latent $h_t$. Defining the distillation target on $h_t$, before quantization and before the
bridge, therefore makes the recipe interface-agnostic. A student trained once can
be paired with either boundary.

\subsection{Distillation Objective}
\label{sec:objective}

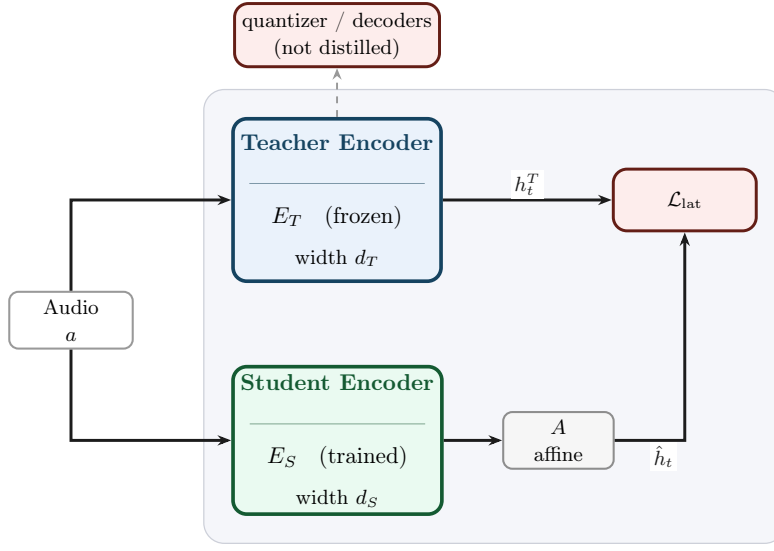
\begin{figure}[t]
  \centering
  \resizebox{0.62\textwidth}{!}{%
    \begin{tikzpicture}[node distance=0pt]
      \node[term] (AUD) at (0,0) {Audio\\[1pt]$a$};
      \node[bigblock, right=1.5cm of AUD, anchor=west, yshift=1.85cm] (TE) {
        {\normalsize\bfseries\color{cEB}Teacher Encoder}\\[3pt]
        {\color{cEB!55}\rule{2.7cm}{0.35pt}}\\[5pt]
        {\normalsize $E_T$ \; (frozen)}\\[6pt]
        {\small width $d_T$}
      };
      \node[smallblock, right=1.5cm of AUD, anchor=west, yshift=-1.85cm] (SE) {
        {\normalsize\bfseries\color{cTB}Student Encoder}\\[3pt]
        {\color{cTB!55}\rule{2.7cm}{0.35pt}}\\[5pt]
        {\normalsize $E_S$ \; (trained)}\\[6pt]
        {\small width $d_S$}
      };
      \node[opell, right=0.9cm of SE] (ADP) {$A$\\[1pt]affine};
      \node[auxblock, right=2.6cm of TE] (LOSS) {$\mathcal{L}_{\text{lat}}$};
      \node[auxblock, above=0.75cm of TE, minimum width=3.1cm] (HEADS)
        {\small quantizer / decoders\\[1pt](not distilled)};
      \begin{pgfonlayer}{background}
        \node[fill=cPanelBg, draw=cPanelBd, rounded corners=9pt, line width=0.55pt,
          fit=(TE)(SE)(ADP)(LOSS), inner sep=12pt] {};
      \end{pgfonlayer}
      \draw[fwd] (AUD) |- (TE.west);
      \draw[fwd] (AUD) |- (SE.west);
      \draw[fwd] (TE.east) -- node[lbl, above]{$h^T_t$} (LOSS.west |- TE.east);
      \draw[fwd] (SE.east) -- (ADP.west);
      \draw[fwd] (ADP.east) -| node[lbl, below, pos=0.35]{$\hat{h}_t$} (LOSS.south);
      \draw[fwdlight] (TE.north) -- (HEADS.south);
    \end{tikzpicture}%
  }
  \caption{Latent-space encoder distillation. The teacher is frozen and only its
  encoder is used; the quantizer and the stage-0 transcription/reconstruction
  decoders play no role in the objective. An affine layer $A$ maps the
  student's width $d_S$ onto the teacher's $d_T$ so the two latents are
  directly comparable.}
  \label{fig:distill}
\end{figure}

The goal is an on-device student tokenizer that reproduces the behavior of the
large teacher tokenizer. We train the encoder portion of the student
model \emph{only}, and we supervise it against the teacher's per-frame latent
rather than against any discrete or probabilistic output.
Figure~\ref{fig:distill} summarizes the arrangement.

Let $a$ denote an input utterance, $E_T$ the frozen teacher encoder producing
$h^T_t \in \mathbb{R}^{d_T}$, and $E_S$ the student encoder producing
$h^S_t \in \mathbb{R}^{d_S}$. Because the teacher and student hidden
dimensions do not match, the student encoder carries an affine layer
$A: \mathbb{R}^{d_S} \to \mathbb{R}^{d_T}$, playing the role that the regressor
plays in FitNets~\citep{romero2015fitnetshintsdeepnets}:
\begin{align}
  h^T_t &= \big[E_T(a)\big]_t , \\
  \hat{h}_t &= A\big(\big[E_S(a)\big]_t\big) .
\end{align}
The training objective is the masked squared error between the teacher and
student latent representations, summed over channels and averaged over frames:
\begin{align}
  \mathcal{L}_{\text{lat}} &= \frac{1}{T} \sum_{t=1}^{T} m_t \sum_{i=1}^{d_T}
      \big( h^T_{t,i} - \hat{h}_{t,i} \big)^{2} ,
  \label{eq:lat}
\end{align}
where $m_t = 1$ on valid frames and $0$ on padded ones. The average is taken over
all $T$ positions rather than over the valid ones alone; Appendix~\ref{app:objective}
gives the batched form and the consequences of that choice.

Only $E_S$ and $A$ receive gradients, so what
distillation produces is an encoder rather than a complete tokenizer. The
teacher's quantizer, and in the stage-0 case its transcription and reconstruction
decoders, are never distilled; the student inherits the quantizer (discrete interface) or the bridge (continuous
interface) from the teacher-side configuration. This keeps the student encoder small, while the full training graph, teacher
decoders included, is several times larger.

Matching the continuous latent under
Eq.~\ref{eq:lat} sidesteps the question of whether the student's codebook
assignments agree with the teacher's. That objective is harder to optimize, because it requires differentiating
through the $\arg\min$ of quantization~\citep{oord2018neuraldiscreterepresentationlearning,mentzer2023finitescalarquantizationvqvae}, and it
is unnecessary, since downstream accuracy depends on the latent the model actually
consumes; Appendix~\ref{app:sufficiency} makes that precise. Because the target is a real-valued vector rather than a distribution
over classes, the objective uses no temperature, no softmax and no label
smoothing.

The squared difference is summed over the $d_T$ channels
and averaged over frames, so the loss scales with the teacher's width; with no
second term in the objective, that scale is absorbed entirely by the learning
rate.

\subsection{Student Encoder}
\label{sec:student}

The student is not a uniformly scaled-down teacher. Both encoders are built from
the same block type, a convolution-augmented self-attention block with
macaron-style feed-forward pairs~\citep{gulati2020conformerconvolutionaugmentedtransformerspeech}. The student, however, trades width for depth: it is substantially narrower than
every teacher ($d_S < d_T$), and deeper than the depth-pruned teacher~C though not
than the deepest. Narrowing is the source of the $2.8\times$ parameter
reduction, since block cost grows quadratically in width but only linearly in
depth~\citep{kim2022squeezeformerefficienttransformerautomatic,yao2024zipformerfasterbetterencoder}.

This trade-off does not incur additional computation. The dominant per-frame terms of a block are the macaron feed-forward pair and the
attention projections, which scale as $4dd_{\text{ff}}+4d^2$. Counting these, the student's per-frame multiply--accumulate cost is
$2.8\times$ lower than the teacher's, matching the parameter reduction. The student's improved quality per parameter is therefore not
obtained at the cost of additional depth-driven computation.

The affine layer $A$ is the only structural addition the student carries relative to
a standalone tokenizer of the same shape, and it accounts for well under $1\%$ of the student encoder. It exists to make Eq.~\ref{eq:lat} well-typed, and is either folded into the
downstream interface block or dropped, so it costs nothing at inference.

\subsection{Distillation Stages}
\label{sec:where}

The teacher tokenizer is trained in several stages; two are relevant here.

Stage~0 trains the tokenizer with no language model present, so that the tokenizer
learns to transcribe the audio and to reconstruct it, via two decoder blocks
attached to the encoder output. Stage~1 then trains the tokenizer jointly with the
language model, initializing the tokenizer from the stage-0 pretrained checkpoint
and the language model from its text-pretrained checkpoint.

For both stages, the teacher tokenizer can be separated from its surrounding
graph and used as the frozen target in Eq.~\ref{eq:lat}. The result is a smaller
tokenizer that drops into the teacher's place, at teacher-comparable word error
rate and without any fine-tuning of the student
(Section~\ref{sec:experiments}).

The two choices are not equivalent in what they commit to. A stage-0 student
targets a language-model-agnostic representation and can subsequently be paired
with, or jointly trained against, any downstream model. A stage-1 student
targets a representation the language model has already co-adapted to, which
removes the risk of a distribution mismatch at the boundary but ties the student
to that model. We report both.

\section{Experiments and Results}
\label{sec:experiments}

\subsection{Data}

Distillation requires no labels, since the objective in Eq.~\ref{eq:lat} is
defined against the teacher's own activations. Students are therefore trained on
unlabeled audio drawn from the same multilingual mixture used for teacher
pretraining, with the teacher run in inference mode on each batch. Audio is
16\,kHz and the encoder emits one latent per 80\,ms.

Word error rate is reported on LibriSpeech dev-clean and
dev-other~\citep{7178964}, TED-LIUM~\citep{Hernandez_2018},
VoxPopuli~\citep{wang2021voxpopulilargescalemultilingualspeech}, AMI~\citep{inbook}, and
Earnings-22~\citep{delrio2022earnings22practicalbenchmarkaccents} for stage-0 runs, covering read,
prepared, parliamentary, meeting, and accented long-form speech. Stage-1 runs
use the evaluation suite configured in the joint training job, which replaces
the meeting and long-form sets with the aggregate multilingual evaluators
MLS~\citep{Pratap_2020}, Common Voice~\citep{ardila2020commonvoicemassivelymultilingualspeech}, and
FLEURS~\citep{conneau2022fleursfewshotlearningevaluation}. The two suites differ, so stage-0 and stage-1
numbers are tabulated separately and should not be compared across tables.

\subsection{Distillation Setup}

All students are trained for 500k steps at a global batch size of 1280 with AdamW, a peak learning rate of $10^{-3}$ under an
inverse-square-root schedule with 10k warmup steps, gradient-norm clipping at 1.0,
and an exponential moving average of the student weights with decay $0.9999$
warmed up over 30k steps; the EMA weights are the ones evaluated.
Training is in \texttt{float32}. The teacher is frozen throughout, enforced by a
learner rule marking its parameters non-trainable, and only its encoder is
instantiated; the quantizer is loaded but receives no gradient, and the two
decoders are excluded entirely.

We evaluate three stage-0 teachers, labeled by encoder architecture and token
interface: \textbf{A} (Conformer, discrete), \textbf{B} (Transformer, continuous) and
\textbf{C} (pruned Conformer, continuous). B replaces A's Conformer
blocks with a standard Transformer stack, and C is a depth-pruned variant of A's
Conformer with half as many blocks, differing from it in training as well as in
size. The three encoders
therefore differ in capacity as well as architecture, so each teacher defines its
own compression ratio against the common student capacity. We further evaluate
three stage-1 joint models, \textbf{J1}--\textbf{J3}: J1 is built on teacher B, J2
and J3 on teacher C's pruned Conformer, and J3 restricts which parameters receive
gradients during joint training.

Each teacher yields one distilled student, giving six teacher--student pairs. Since
an encoder produces no transcript by itself, WER is measured through the full
recognition stack: the teacher's 262M-parameter transcription decoder for stage-0
pairs, and the 2.8B-parameter joint language model for stage-1 pairs. The encoder
is the only component that differs between a teacher and its student, and no
student is fine-tuned. As controls we train two students of the same capacity
without a teacher, although both inherit four initialized sub-blocks from a
distilled model, which makes them conservative rather than clean from-scratch
baselines.

\subsection{Results on Stage-0 Distillation}

\begin{table}[t]
\centering
\caption{Stage-0 distillation. Word error rate (\%), lower is better. Recognition is performed by the teacher's 262M-parameter transcription decoder,
which is identical across every row, so only the encoder differs. Each teacher/student pair is separated by the distillation
step only; students are evaluated \emph{without} fine-tuning. The last block gives
same-capacity students trained independently rather than distilled.}
\label{tab:stage0}
\resizebox{\textwidth}{!}{%
\begin{tabular}{@{}llcccccc|c@{}}
\toprule
\textbf{Teacher} & \textbf{Model} & \textbf{lib-clean} & \textbf{lib-other} &
\textbf{TED-LIUM} & \textbf{VoxPopuli} & \textbf{AMI} & \textbf{Earnings-22} &
\textbf{Avg.} \\
\midrule
\multirow{2}{*}{A (Conformer, discrete)}
  & teacher           & 2.61 & 6.54 & 7.25 & 11.73 & 30.05 & 29.63 & 14.64 \\
  & distilled        & 2.73 & 6.65 & 7.37 & \textbf{11.65} & \textbf{29.87} & 29.90 & 14.70 \\
\cmidrule(lr){1-9}
\multirow{2}{*}{B (Transformer, continuous)}
  & teacher           & 1.92 & 4.68 & 5.64 & 11.06 & 23.89 & 22.87 & 11.68 \\
  & distilled        & \textbf{1.91} & 4.82 & 6.10 & \textbf{11.05} & \textbf{23.54} & 23.99 & 11.90 \\
\cmidrule(lr){1-9}
\multirow{2}{*}{C (Conformer, continuous)}
  & teacher           & 1.85 & 4.49 & 5.79 & 11.02 & 24.36 & 23.98 & 11.91 \\
  & distilled        & 1.88 & 4.80 & \textbf{5.77} & 11.09 & \textbf{24.05} & 24.45 & 12.01 \\
\midrule
\multicolumn{2}{@{}l}{\textit{no distillation, same student capacity}} & & & & & & & \\
\multicolumn{2}{@{}l}{Student capacity, independently trained}
                     & 2.05 & 5.14 & 6.62 & 11.19 & 24.14 & 25.04 & 12.36 \\
\multicolumn{2}{@{}l}{Student capacity, frozen decoder}
                     & 1.92 & 4.81 & 6.10 & 11.06 & 24.08 & 25.90 & 12.31 \\
\bottomrule
\end{tabular}%
}
\end{table}

Table~\ref{tab:stage0} reports the three stage-0 teacher/student pairs. Across all three, the distilled student tracks its teacher to within $1.9\%$
relative WER: $+0.4\%$ for teacher A, $+1.9\%$ for teacher B and
$+0.8\%$ for teacher~C (absolute $+0.06$, $+0.22$, $+0.10$ WER). On several individual sets the student is
\emph{better} than its teacher: AMI in all three pairs and VoxPopuli in two. This
is consistent with a small regularization effect from regressing a smoothed
target rather than optimizing the recognition objective
directly~\citep{yuan2021revisitingknowledgedistillationlabel}.

\subsection{Results on Stage-1 Distillation}

\begin{table}[t]
\centering
\caption{Stage-1 distillation, where the teacher has been jointly trained with the
language model. Word error rate (\%), lower is better. Recognition is performed by the 2.8B-parameter joint language model, identical
across every row, so only the encoder differs. The evaluation
suite is the one configured in the joint training job and differs from
Table~\ref{tab:stage0}.}
\label{tab:stage1}
\resizebox{\textwidth}{!}{%
\begin{tabular}{@{}llcccccc|c@{}}
\toprule
\textbf{Joint model} & \textbf{Model} & \textbf{lib-clean} & \textbf{lib-other} &
\textbf{VoxPopuli} & \textbf{MLS} & \textbf{CommonVoice} & \textbf{FLEURS} &
\textbf{Avg.} \\
\midrule
\multirow{2}{*}{J1 (Transformer)}
  & teacher           & 2.04 & 5.07 & 8.30 & 10.50 & 12.52 & 6.84 & 7.54 \\
  & distilled        & 2.07 & 5.11 & 8.42 & 10.54 & 12.98 & \textbf{6.80} & 7.65 \\
\cmidrule(lr){1-9}
\multirow{2}{*}{J2 (Conformer)}
  & teacher           & 3.38 & 8.50 & 9.09 & 19.77 & 21.49 & 11.76 & 12.33 \\
  & distilled        & \textbf{3.27} & \textbf{8.13} & \textbf{8.94} & \textbf{18.82} & \textbf{20.14} & \textbf{10.76} & \textbf{11.68} \\
\cmidrule(lr){1-9}
\multirow{2}{*}{J3 (Conformer, restr.\ grad.)}
  & teacher           & 2.37 & 5.45 & 8.49 & 14.25 & 13.75 & 7.74 & 8.67 \\
  & distilled        & 2.50 & 5.96 & 8.64 & 15.63 & 14.82 & 8.46 & 9.34 \\
\bottomrule
\end{tabular}%
}
\end{table}

Table~\ref{tab:stage1} repeats the experiment where the teacher has already
co-adapted to a language model. The outcome is less uniform than at stage 0. For J1 the student is within $1.5\%$ relative WER. For J2
the student is \emph{better} than its teacher on every test set,
$11.68$ vs.\ $12.33$ average---a 5.3\% relative improvement from a model with a
third of the encoder parameters. For J3 the student is $7.7\%$ relative worse, the largest degradation we observe
anywhere.
\subsection{Ablation Studies}

\subsubsection{Distillation vs.\ Independent Training}

The comparison that motivates the recipe is distillation against
training the small tokenizer independently. The last block of
Table~\ref{tab:stage0} gives two such controls at identical capacity: a
student-capacity model trained independently reaches 12.36 average WER and a
frozen-decoder variant 12.31, against 11.90 for the distilled student of the
same size. Distillation is worth $3.9\%$ relative at fixed capacity ($0.46$ WER absolute).

The gap is widest where a small model is most starved for supervision:
Earnings-22 ($23.99$ vs.\ $25.04$) and TED-LIUM ($6.10$ vs.\ $6.62$). It is nearly
closed on clean read speech ($1.91$ vs.\ $2.05$). Distillation therefore yields robustness on the hard
conditions rather than accuracy on the easy ones.

\subsubsection{Effect of Teacher Quality}

Across all six pairs, the strongest predictor of student WER is teacher WER, not
the distillation configuration. This runs counter to the capacity-gap findings
of~\citep{cho2019efficacyknowledgedistillation,mirzadeh2019improvedknowledgedistillationteacher}, where a stronger teacher can transfer
less well to a small student. Teacher A in Table~\ref{tab:stage0} is $25\%$ relative worse than teacher B,
and its student inherits nearly the whole gap ($14.70$ vs.\ $11.90$, a
difference of 2.80). The relationship holds in stage 1 as well: the weakest
teacher (J2, 12.33) yields the weakest stage-1 student (11.68), and the strongest teacher (J1, 7.54) yields the strongest student (7.65).

The practical implication is that effort spent improving the teacher transfers
almost one-for-one to the student, whereas the distillation step itself is
already close to lossless and leaves little to recover. The choice of teacher
therefore matters more than the details of the distillation procedure.

\subsubsection{Discrete vs.\ Continuous Interfaces}

Because the objective is defined before the quantizer, the same recipe applies
unchanged to both interfaces and transfers equally well: $+0.4\%$ relative for the discrete-interface pair against $+1.9\%$ and $+0.8\%$ for the
continuous ones. The
interface need not be fixed before distilling. The large absolute difference between the discrete- and continuous-interface rows
in Table~\ref{tab:stage0} is a property of the teachers, not of
the distillation: as noted in Section~\ref{sec:objective}, quantization
loss is incurred on both sides of the teacher/student comparison and therefore cancels.

\subsubsection{Effect of the Distillation Stage}

All three stage-0 students degrade by similar amounts ($+0.4\%$ to $+1.9\%$
relative), whereas the three stage-1 students span $-5.3\%$ to $+7.7\%$. We attribute the wider spread to the
teacher having co-adapted to a particular language model: the quality of the
latent geometry being regressed then depends on how well that joint training
went, which varies across the three joint models in a way stage-0 pretraining
does not.

Stage 0 is also the more reusable: its student targets a language-model-agnostic
representation, whereas a stage-1 student is tied to the model it was distilled
under. We therefore default to stage 0, and reach for stage 1 only when the
tokenizer must match an already-trained joint model.

\section{Conclusion}
\label{sec:conclusion}

In this work, we introduced a latent-space distillation recipe for audio
tokenizers, which compresses a tokenizer encoder by $2.8\times$ at teacher-comparable word
error rate. We studied the importance of
each design choice, and showed that supervising the \emph{pre-quantizer
latent}---the last representation the two token interfaces share---is sufficient
to make the recipe portable: the same student serves both the discrete and continuous interfaces, and the same objective applies whether the teacher was pretrained
alone or jointly trained with a language model. Across six teacher--student
pairs, five of six students fall within $1.9\%$ relative WER of their teacher
without fine-tuning, and the distilled student outperforms an independently
trained tokenizer of identical capacity by $3.9\%$ relative. Our ablations further show that student quality
is governed chiefly by teacher quality, that the token interface does not affect
how well distillation transfers, and that stage-0 distillation is both more
predictable and less constraining than stage-1.

The principal result is one of efficiency. The full-size always-on front end could not be accommodated within the memory
and latency budget it must share with the foundation model it feeds; a
$2.8\times$ smaller encoder, at correspondingly lower compute, can be, and latent-space distillation is what closed that gap
without requiring anything downstream of the encoder to be retrained. At that
size the encoder operates on-device as the audio front end for system-wide
Dictation, feeding a sparsely activated foundation model under the shared DRAM
budget that IFP creates. The production configuration is tuned separately from the
research configuration reported here and differs from it in several respects;
unless stated
otherwise every number in this paper refers to the research configuration, and all
word error rates are public-benchmark results measured under our research
evaluation protocol.

\bibliographystyle{plainnat}
\bibliography{refs}

%% formal appendices -- arXiv version only (the ICASSP build is over its limit)
% =====================================================================
% Formal appendices for the Apple MLR / arXiv version.
%
% Included only by paper_mlr/main.tex -- NOT by the ICASSP build, which is
% already over its page limit.  Content is ported from
% ../expanded_report/sections/04_method.tex, restated symbolically so that no
% absolute encoder configuration is disclosed (see paper_v2/obfuscate_config.py).
% =====================================================================

\appendix

\section{Formal Description of the Objective}
\label{app:formal}

Section~\ref{sec:objective} states the objective in words. This appendix states it
formally, including the reduction order and the masking convention that a
reproduction needs.

\subsection{Notation}
\label{app:notation}

\begin{table}[htbp]
\centering
\small
\begin{tabular}{@{}l l l@{}}
\toprule
\textbf{Symbol} & \textbf{Meaning} & \textbf{Domain} \\
\midrule
$B$                    & batch size                                        & $1{,}280$ \\
$T$                    & output frames per example (one per 80\,ms)         & variable \\
$L$                    & input filterbank frames (one per 10\,ms)          & $L = 8T$ \\
$d_{\mathrm{fb}}$      & filterbank feature dimension                      & $80$ \\
$d_S$                  & student internal Conformer width                  & $d_S < d_T$ \\
$d_T$                  & teacher output width, and the interface width     & --- \\
$d_{\mathrm{LM}}$      & language-model embedding width                    & --- \\
$\mathbf{X}$           & filterbank features for one example               & $\Rb^{L \times d_{\mathrm{fb}}}$ \\
$\mathbf{s}$           & per-frame segment identifiers ($0 =$ padding)     & $\mathbb{Z}^{B \times T}$ \\
$\mathbf{Y}$           & teacher pre-quantizer latents                     & $\Rb^{B \times T \times d_T}$ \\
$\widehat{\mathbf{Y}}$ & student pre-quantizer latents                     & $\Rb^{B \times T \times d_T}$ \\
$\mathbf{M}$           & validity mask                                     & $\{0,1\}^{B \times T}$ \\
$\enc_T, \enc_S$       & teacher and student encoders                       & --- \\
$\theta_S$             & student encoder parameters                         & --- \\
\bottomrule
\end{tabular}
\caption{Notation. Widths are given symbolically; only their ordering and the
$2.8\times$ parameter ratio are reported (Section~\ref{sec:student}).}
\label{tab:notation}
\end{table}

\subsection{The encoder as a composition of operators}
\label{app:operators}

Both encoders factor into the same stages. Naming them is worthwhile because the
distillation objective attaches at a specific point in the composition.

Given filterbank features $\mathbf{X} \in \Rb^{L \times d_{\mathrm{fb}}}$ computed
at one frame per 10\,ms, we apply a causal convolutional subsampler $\sub$ that
reduces the frame rate by $4\times$, a causal Conformer stack $\conf$ with
speech-context attention, a post-downsample $\down$ that reduces the frame rate by
a further $2\times$, and---in the student only---an affine layer $\aff$. The
student encoder is
\begin{equation}
\label{eq:student-enc}
  \enc_S(\mathbf{X}, \mathbf{s})
  \;=\;
  \aff\bigl(\down(\conf_{n_S}(\sub(\mathbf{X}), \mathbf{s}))\bigr)
  \;\in\; \Rb^{T \times d_T},
  \qquad
  T = \tfrac{L}{8},
\end{equation}
and the teacher encoder is the same composition without the affine layer,
\begin{equation}
\label{eq:teacher-enc}
  \enc_T(\mathbf{X}, \mathbf{s})
  \;=\;
  \down(\conf_{n_T}(\sub(\mathbf{X}), \mathbf{s}))
  \;\in\; \Rb^{T \times d_T}.
\end{equation}
The $8\times$ total temporal reduction is what makes one output vector correspond
to 80\,ms of audio. The two stacks differ in both depth and width. The student is
substantially narrower than every teacher we distill from ($d_S < d_T$); its depth
lies between that of the deepest teacher and that of the depth-pruned one, so the
relation between $n_S$ and $n_T$ is not uniform across pairs.

Downstream, the latent reaches the language model by one of two routes,
\begin{equation}
\label{eq:interfaces}
  \underbrace{\quant\bigl(\enc(\mathbf{X}, \mathbf{s})\bigr) \in \mathbb{Z}^{T}}_{\text{discrete}}
  \qquad\text{or}\qquad
  \underbrace{\bridge\bigl(\enc(\mathbf{X}, \mathbf{s})\bigr) \in \Rb^{T \times d_{\mathrm{LM}}}}_{\text{continuous}},
\end{equation}
where $\quant$ is the vector quantizer and $\bridge$ the linear bridge projection.
Equation~\ref{eq:interfaces} is the whole argument for the choice of target: both
routes take $\enc(\mathbf{X}, \mathbf{s})$ as their only argument, so an objective
on $\enc$'s output is simultaneously an objective for both.

\subsection{The masked objective}
\label{app:objective}

Write $\mathbf{Y} = \enc_T(\mathbf{X}, \mathbf{s})$ and
$\widehat{\mathbf{Y}} = \enc_S(\mathbf{X}, \mathbf{s})$, batched to
$\Rb^{B \times T \times d_T}$. Note that $\widehat{\mathbf{Y}}$ already carries the
teacher's width: the affine layer lives \emph{inside} $\enc_S$, so the two latents are
directly comparable with no external projection.

Padding is handled through the segment identifiers rather than a separate padding
tensor. The validity mask is
\begin{equation}
\label{eq:mask}
  M_{b,t}
  \;=\;
  \begin{cases}
    0 & \text{if } s_{b,t} = 0 \quad (\text{padded frame}), \\
    1 & \text{otherwise},
  \end{cases}
\end{equation}
and the objective is a masked squared-error latent-matching loss,
\begin{equation}
\label{eq:loss-full}
  \Lmatch(\theta_S)
  \;=\;
  \frac{1}{BT}
  \sum_{b=1}^{B} \sum_{t=1}^{T} \sum_{k=1}^{d_T}
  \Bigl( M_{b,t}\bigl( Y_{b,t,k} - \widehat{Y}_{b,t,k} \bigr) \Bigr)^{2},
\end{equation}
minimized over $\theta_S$ only, with the teacher held fixed. In words: take the
per-frame difference between teacher and student latents, zero it on padded
frames, square it, sum over the $d_T$ channels, and average over all $BT$
positions.

Two properties of Equation~\ref{eq:loss-full} are load-bearing and easy to get
wrong when reimplementing.

The channel reduction is a sum rather than a mean: the squared difference is
summed over the $d_T$ channels, so the loss scales with the teacher's width. With
no other term in the objective that scale is absorbed entirely by the learning
rate, but it matters if Equation~\ref{eq:loss-full} is ever compared across widths
or combined with a second term at a fixed weight. A multi-term successor would be
better served by normalizing each term's gradient to a common scale, as the loss
balancer of \citet{défossez2022highfidelityneuralaudio} does, so that the weights
denote gradient fractions rather than inheriting the arbitrary magnitude of each
loss.

The position reduction, in turn, divides by $BT$ rather than by
$\sum_{b,t} M_{b,t}$, so padded positions contribute exactly zero to the numerator
yet are still counted in the denominator. Writing $\rho = \bigl(\sum_{b,t} M_{b,t}\bigr)/BT$ for the
fraction of valid frames in a batch,
\begin{equation}
\label{eq:normvariant}
  \Lmatch \;=\; \rho \cdot \Lnorm,
  \qquad
  \Lnorm \;=\; \frac{1}{\textstyle\sum_{b,t} M_{b,t}}
  \sum_{b,t,k} \Bigl( M_{b,t}\bigl( Y_{b,t,k} - \widehat{Y}_{b,t,k} \bigr) \Bigr)^{2},
\end{equation}
so the gradient is scaled by a data-dependent factor $\rho$ that varies with the
batch's padding ratio. Under a fixed batching strategy this is a constant and
harmless; under variable-length bucketing it mildly upweights densely packed
batches. We did not isolate the effect of using $\Lnorm$ instead.

An $\ell_1$ variant, obtained by replacing the square in
Equation~\ref{eq:loss-full} with an absolute value, is a reasonable alternative:
it is less sensitive to occasional large per-frame deviations, which in a
distillation setting are often frames the teacher itself is uncertain about. All
results reported here use Equation~\ref{eq:loss-full}; we did not run the
comparison.

\subsection{Why matching the latent suffices}
\label{app:sufficiency}

The portability claim can be stated precisely. Let $f$ denote any downstream map
applied to the latent---either $\quant$ followed by an embedding lookup, or
$\bridge$ followed by the language model. Suppose the student matches the teacher
to within $\varepsilon$ at some frame, $\lVert \hat{h} - h \rVert_2 \le
\varepsilon$. Two regimes follow, one for each interface.

The continuous interface degrades smoothly and boundedly. Because the bridge is
linear,
\begin{equation}
\label{eq:cont-bound}
  \bigl\lVert \bridge(\hat{h}) - \bridge(h) \bigr\rVert_2
  \;=\; \bigl\lVert W_{\text{bridge}}(\hat{h} - h) \bigr\rVert_2
  \;\le\; \lVert W_{\text{bridge}} \rVert_2 \, \varepsilon ,
\end{equation}
with $\lVert \cdot \rVert_2$ the spectral norm. The perturbation entering the
language model is therefore controlled by the latent error, uniformly over inputs.

Discrete tokens, by contrast, are exactly invariant below a margin. Letting $c^\star$
denote the codebook index nearest to $h$, define the margin
\begin{equation}
\label{eq:margin}
  \gamma(h) \;=\; \min_{c \neq c^\star}
  \bigl( \lVert h - e_c \rVert_2 - \lVert h - e_{c^\star} \rVert_2 \bigr) \;>\; 0 .
\end{equation}
For any $c$ we have $\lVert \hat{h} - e_c \rVert_2 \ge \lVert h - e_c \rVert_2 -
\varepsilon$ and $\lVert \hat{h} - e_{c^\star} \rVert_2 \le \lVert h - e_{c^\star}
\rVert_2 + \varepsilon$, so $\hat{h}$ still selects $c^\star$ whenever
$\lVert h - e_c \rVert_2 - \varepsilon > \lVert h - e_{c^\star} \rVert_2 +
\varepsilon$ for every $c \neq c^\star$, that is whenever
\begin{equation}
\label{eq:disc-condition}
  \varepsilon \;<\; \tfrac{1}{2}\,\gamma(h) .
\end{equation}
Below that threshold the emitted token is \emph{identical}, and the language model
cannot distinguish student from teacher at all.

Equations~\ref{eq:cont-bound} and~\ref{eq:disc-condition} are sufficient
conditions, not an account of the measured differences: $\varepsilon$ is not
uniform across frames, and Equation~\ref{eq:disc-condition} fails on frames that
sit near a Voronoi boundary---precisely the frames where the teacher's own
assignment is least stable. They do explain why one objective placed on $h$ serves
both interfaces without modification, and why neither interface transfers
systematically better than the other (Section~\ref{sec:experiments}).

\subsection{The affine layer}
\label{app:affine}

The student's Conformer is narrower than the teacher's. A single affine map closes
the gap:
\begin{equation}
\label{eq:affine}
  \aff(\mathbf{H}) \;=\; \mathbf{H}\mathbf{W} + \mathbf{b},
  \qquad
  \mathbf{W} \in \Rb^{d_S \times d_T},\;
  \mathbf{b} \in \Rb^{d_T},
\end{equation}
appended inside the student encoder so that its declared output width is the
teacher's. The construction is conditional: $\aff$ exists only if $d_S \neq
d_T$, in which case the student's advertised output width is overwritten with
$d_T$. If the widths already match, no affine layer is created and
Equation~\ref{eq:loss-full} compares outputs directly.

The affine layer costs $d_S \cdot d_T$ parameters, well under $1\%$ of the student
encoder. It also absorbs a change of basis: because
Equation~\ref{eq:loss-full} compares in the teacher's coordinate system, without
$\aff$ the student's final Conformer layer would have to both compute features
and change basis.

\subsection{What is frozen and what is discarded}
\label{app:frozen}

The teacher is frozen by an explicit learner rule marking its parameters
non-updatable. It is additionally invoked with training mode disabled, so its
normalization uses inference statistics and no dropout is applied, and its
gradients are never computed. Its quantizer is retained but unused during
training: it is loaded, because the evaluation path needs it when the
pre-quantizer flag is not set, but it receives no gradient.

The student's decoders are discarded before training begins, leaving $\enc_S$ and
nothing else in the student graph. This is deliberate. The teacher's transcription
decoder ($262$M) and its reconstruction decoder are together larger than its
encoder, neither runs on-device, and every student parameter spent reproducing
them is a parameter not spent on the encoder that ships.

\section{Training Configuration}
\label{app:optim}

The trainer is cloned from the teacher's own experiment configuration with the
model and learner replaced, so the data pipeline, sharding, and evaluation harness
are inherited rather than reimplemented.

\begin{table}[htbp]
\centering
\small
\begin{tabular}{@{}l l@{}}
\toprule
\textbf{Setting} & \textbf{Value} \\
\midrule
Optimizer                        & AdamW \\
Peak learning rate $\eta_{\max}$ & $1 \times 10^{-3}$ \\
Schedule                         & inverse-square-root decay, \\
                                 & scale $= \eta_{\max}\sqrt{T_{\mathrm{warm}}}$ \\
Warmup steps $T_{\mathrm{warm}}$ & $10{,}000$ \\
Gradient clipping                & global norm $1.0$ \\
EMA decay                        & $0.9999$, warmed up over $30{,}000$ steps \\
Batch size $B$                   & $1{,}280$ \\
Max steps                        & $500{,}000$ \\
Precision                        & \texttt{float32} \\
Sample-rate constraint           & teacher and student must match (asserted) \\
\bottomrule
\end{tabular}
\caption{Distillation training configuration, identical for every run reported.}
\label{tab:optim}
\end{table}

An exponential moving average of the student weights is maintained and is what
downstream consumers load. Evaluating the raw weights rather than the EMA weights
will not reproduce the reported numbers.

\section{The Recipe in Full}
\label{app:recipe}

Algorithm~\ref{alg:recipe} states the procedure end to end.

\begin{algorithm}[htbp]
\caption{Latent-space encoder distillation.}
\label{alg:recipe}
\begin{algorithmic}[1]
\Require teacher configuration and checkpoint $\mathcal{K}_T$;
         student configuration; steps $N$; batch size $B$; EMA decay $\beta$
\Ensure  distilled student encoder $\enc_S$

\Statex \textbf{\textit{Setup}}
\If{$\mathcal{K}_T$ is a joint tokenizer$+$language-model checkpoint}
  \State $\mathcal{K}_T \gets \textsc{ExtractTokenizer}(\mathcal{K}_T)$
         \Comment{scope surgery}
\EndIf
\State instantiate $\enc_T$; load $\mathcal{K}_T$ into the teacher scope
\State set the pre-quantizer flag on $\enc_T$
       \Comment{read the latent of Eq.~\ref{eq:interfaces}}
\State mark every teacher parameter \textsc{NoUpdate}; disable training mode
       \Comment{App.~\ref{app:frozen}}
\State instantiate $\enc_S$; empty its decoder list
\If{$d_S \neq d_T$}
  \State append $\aff : \Rb^{d_S} \!\to\! \Rb^{d_T}$ inside $\enc_S$
         \Comment{Eq.~\ref{eq:affine}}
\EndIf
\State $\bar{\theta}_S \gets \theta_S$ \Comment{EMA accumulator}

\Statex \textbf{\textit{Training}}
\For{$i = 1$ \textbf{to} $N$}
  \State draw an unlabeled batch $(\mathbf{X}, \mathbf{s})$ of size $B$
  \State $\mathbf{Y} \gets \enc_T(\mathbf{X}, \mathbf{s})$
         \Comment{frozen; no gradient}
  \State $\widehat{\mathbf{Y}} \gets \enc_S(\mathbf{X}, \mathbf{s})$
  \State $M_{b,t} \gets \mathds{1}\!\left[ s_{b,t} \neq 0 \right]$
         \Comment{Eq.~\ref{eq:mask}}
  \State $\Lmatch \gets \dfrac{1}{BT} \sum_{b,t,k}
         \bigl( M_{b,t} ( Y_{b,t,k} - \widehat{Y}_{b,t,k} ) \bigr)^{2}$
         \Comment{Eq.~\ref{eq:loss-full}}
  \State $g \gets \textsc{ClipGlobalNorm}\bigl( \nabla_{\theta_S} \Lmatch,\, 1.0 \bigr)$
  \State $\theta_S \gets \textsc{AdamW}(\theta_S,\, g,\, \eta_i)$
         \Comment{$\eta_i$ per Tab.~\ref{tab:optim}}
  \State $\bar{\theta}_S \gets \beta \bar{\theta}_S + (1 - \beta) \theta_S$
\EndFor

\State \Return $\bar{\theta}_S$
       \Comment{the EMA weights, not $\theta_S$}
\end{algorithmic}
\end{algorithm}

Changing the teacher requires editing exactly two values: the teacher
configuration name and the teacher checkpoint path. That is the property which
made the recipe reusable across six teacher--student pairs and three joint models.

\applefootnote{ \textcolor{textgray}{\sffamily Apple and the Apple logo are trademarks of Apple Inc., registered in the U.S. and other countries and regions.}}

\end{document}